\documentclass[sigconf]{acmart}

\copyrightyear{2025}
\acmYear{2025}
\setcopyright{acmlicensed}\acmConference[WWW Companion '25]{Companion Proceedings of the ACM Web Conference 2025}{April 28-May 2, 2025}{Sydney, NSW, Australia}
\acmBooktitle{Companion Proceedings of the ACM Web Conference 2025 (WWW Companion '25), April 28-May 2, 2025, Sydney, NSW, Australia}
\acmDOI{10.1145/3701716.3735081}
\acmISBN{979-8-4007-1331-6/2025/04}

\usepackage{xspace}
\usepackage{booktabs}
\usepackage{textcomp}
\usepackage{amsmath}
\usepackage{hyperref}
\usepackage{subcaption}
\newcommand{\mName}{DisCo\xspace}

\begin{document}

\title{Personalized Diffusion Model Reshapes Cold-Start Bundle Recommendation}

\author{Tuan-Nghia Bui}
\email{21020364@vnu.edu.vn}
\affiliation{
  \institution{VNU University of Engineering and Technology}
  \city{Hanoi}
  \country{Vietnam}
}

\author{Huy-Son Nguyen}
\email{huyson@vnu.edu.vn}
\affiliation{%
  \institution{VNU University of Engineering and Technology}
  \city{Hanoi}
  \country{Vietnam}
}

\author{Cam-Van Thi Nguyen}
\email{vanntc@vnu.edu.vn}
\affiliation{%
  \institution{VNU University of Engineering and Technology}
  \city{Hanoi}
  \country{Vietnam}
}

\author{Hoang-Quynh Le}
\email{lhquynh@vnu.edu.vn}
\affiliation{%
  \institution{VNU University of Engineering and Technology}
  \city{Hanoi}
  \country{Vietnam}
}

\author{Duc-Trong Le}
\email{trongld@vnu.edu.vn}
\affiliation{%
  \institution{VNU University of Engineering and Technology}
  \city{Hanoi}
  \country{Vietnam}
}

\renewcommand{\shortauthors}{Tuan-Nghia Bui, Huy-Son Nguyen, Cam-Van Thi Nguyen, Hoang-Quynh Le, \& Duc-Trong Le}

\begin{abstract}
Bundle recommendation aims to recommend a set of items to each user. However, the sparser interactions between users and bundles raise a big challenge, especially in cold-start scenarios. Traditional collaborative filtering methods do not work well for this kind of problem because these models rely on interactions to update the latent embedding, which is hard to work in a cold-start setting. 
We propose a new approach (\mName{}), which relies on a personalized \underline{Di}ffu\underline{s}ion backbone, enhanced by disentangled aspects for the user's interest, to generate a bundle in distribution space for each user to tackle the \underline{co}ld-start challenge. 
During the training phase, \mName{} adjusts an additional objective loss term to avoid bias, a prevalent issue while using the generative model for top-$K$ recommendation purposes.
Our empirical experiments show that \mName{} outperforms five comparative baselines by a large margin on three real-world datasets. 
Thereby, this study devises a promising framework and essential viewpoints in cold-start recommendation.   
Our materials for reproducibility are available at: \href{https://github.com/bt-nghia/DisCo}{https://github.com/bt-nghia/\mName{}}.
\end{abstract}


\maketitle

\section{Introduction}

In recent years, bundle recommendation have been making many progresses through matrix-factorization~\cite{10.5555/1795114.1795167} methods and graph-based techniques~\cite{ma2022crosscbr, ma2024multicbr, jeon2024cold,nguyen2024bundle}. 
They utilize complex approaches to explore the user preference from multi-types of interactions and leverage Bayesian Personalized Ranking loss~\cite{10.5555/1795114.1795167} to discriminate the true negative sampled bundles through updating the latent embedding representation.
These latent collaborative methods~\cite{10.5555/1795114.1795167, ma2022crosscbr, ma2024multicbr,nguyen2024bundle} perform well in warm-start scenarios, where all users and bundles have historical interactions. However, they overlook a common real-world setting in recommendation systems: cold-start scenarios.
Regardless the success of latent-based methods on warm settings, their efficacy vanishes in cold settings, where the latent embeddings of users and bundles stay nearly unchanged during training and inference. 
The challenge of cold bundle recommendation is even more pronounced than that of cold item recommendation due to the increased sparsity of the interaction matrix, making it significantly more difficult to extract collaborative signals. \citeauthor{jeon2024cold}~\cite{jeon2024cold} investigates the weakness of cold-item recommendation methods on bundle variants and adapt~\cite{ma2022crosscbr} backbone with curriculum learning to tackle the ineffectiveness of these models but still struggle to match the performance of warm configuration.
On the other hand, adapting generative models~\cite{vaswani2017attention, ho2020denoising} can represent cold objects more meaningful, avoiding using unchanged initialized vectors to represent cold objects in recommendation
, especially in bundle recommendation where any bundle can be factorized into items.\\ 
In this paper, we propose \textbf{\mName{}} (\textbf{Di}ffusion Model Re\textbf{s}hapes \textbf{Co}ld-Start Bundle Recommendation) to tackle the common error in latent-based method in cold-start setting, \mName{} leverage Diffusion backbone to generate a probabilistic bundle in distribution space, the reconstruct process is guided by personalized signal enhanced by disentangled self-attention module where each aspect of user's interest is differentiated. During the training phase our model using an additional objective loss term to avoid bias, the common issue while using generative model for recommendation purposes.


\section{Related Works}
\textbf{Bundle Recommendation.} 
State-of-the-art works incorporate bipartite graph structure~\cite{ma2022crosscbr, ma2024multicbr}, 
or item-level high-order interconnections~\cite{nguyen2024bundle,bui2024bridge} 
to enhance the latent representation of users and bundles. However, there is limited research on the performance of these models in cold-start settings. 
CoHeat~\cite{jeon2024cold} is the only work that adopts  ~\cite{ma2022crosscbr} backbone and curriculum learning to transfer the preference from item interest to boost the cold-bundle recommendation, but is still limited by using the traditional CF framework.\\
\textbf{Cold-Start Recommendation.} A long-standing task in the field of recommender systems, existing cold-start recommendation's main interest is to suggest cold items to users or items to users who do not have any or few interactions. Existing works are divided into dropout-based~\cite{volkovs2017dropoutnet} , generative methods~\cite{chen2022generative, sohn2015learning}, replacing cold object by multi-modal feature~\cite{du2020learn}. 
The common of these aforementioned methods is that they avoid using the latent initialized embeddings of cold item directly due to optimizing cold embedding methods are ineffective. Collaborative methods which leverage latent representations often fail in cold-setting~\cite{zhu2020recommendation, chen2022generative}. These methods focus on cold user-item recommendation where the sparsity of interaction matrix is denser than user-bundle ones, which makes it even more challenging. To the best of our knowledge, \mName{} is the first architecture that leveraging distribution space instead of normal cold latent feature as previous works to tackle the prominent issues in cold-start bundle recommendation.


\section{Methodology}

\begin{figure}[t]
\centering
\includegraphics[width=0.47\textwidth]{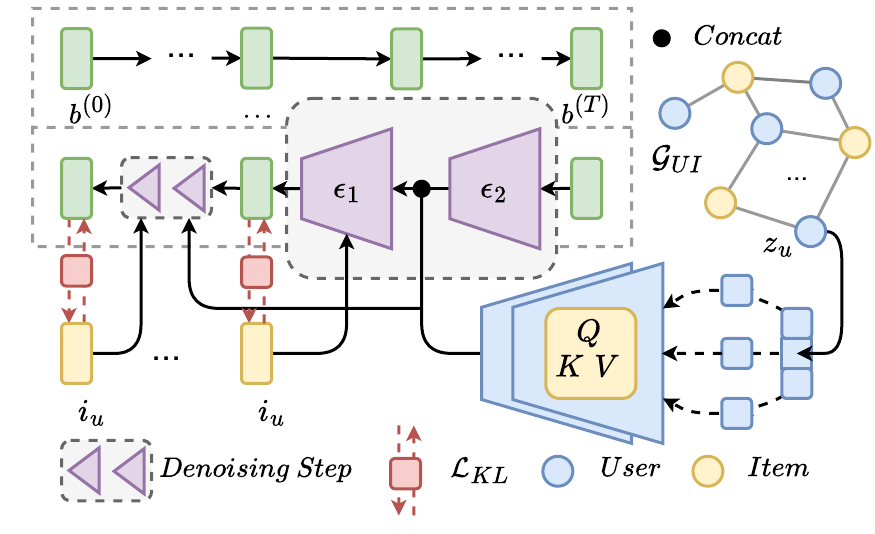}
\vspace{-15px}
\caption{The end-to-end architecture of \mName{}}
\Description{Overall Architecture of Our Model}
\label{over_fig}
\vspace{-15px}
\end{figure}

As shown in Fig~\ref{over_fig}, \mName{} consists of three components: Personalized Conditioning, Guided Diffusion Model, and Optimization. 
\\
\textbf{Problem Formulation.} In this work, we only focus on recommending completely cold bundles for users.
Firstly, let $\mathcal{U}$, $\mathcal{B}$, $\mathcal{I}$ be the sets of all users, bundles and items, with the cold-bundle recommendation, 
the observed interactions of user-bundle, user-item and bundle-item affiliations are defined respectively as three binary matrices ${X} \in \{0, 1\}^{|\mathcal{U}| \times |\mathcal{B}|}$, ${Y} \in \{0, 1\}^{|\mathcal{U}| \times |\mathcal{I}|}$, ${Z} \in \{0, 1\}^{|\mathcal{B}| \times |\mathcal{I}|}$, where $1$ indicates an observed interaction, $0$ for otherwise.
The objective of our approach is to accurately recommend unseen bundles across the bundle set $\mathcal{B}$ (includes warm bundles) to each user $u \in \mathcal{U}$.
At the beginning of the training phase, user and item embeddings are randomly initialized with $d$-dimensions, where $z_u, z_i \in \mathbb{R}^d$ represent a latent feature of user $u$ and item $i$. Besides, a user-item graph $\mathcal{G}_{UI}$ is constructed rely on user-item interaction in $Y$.


\subsection{Personalized Conditioning}
To generate a personalized bundle for different users, we aim to leverage user latent features and historical user-item interactions to guide the generative process. We first enhance user representation using two modules namely: Graph Propagation and Disentangled Self-Attention to obtain personalized latent representation of user.\\
\textbf{Graph Propagation.} \mName{} applies LightGCN~\cite{he2020lightgcn} on User-Item graph ($\mathcal{G}_{UI}$) to propagate high-order collaborative signals as:
\begin{equation}
\begin{split}
    z_u^{l} = \sum_{i \in \mathbb{N}_u} \frac{1}{\sqrt{|\mathbb{N}_u|}\sqrt{|\mathbb{N}_i|}} z_i^{l-1},~~~~
    z_i^{l} = \sum_{u \in \mathbb{N}_i} \frac{1}{\sqrt{|\mathbb{N}_u|}\sqrt{|\mathbb{N}_i|}} z_u^{l-1},
\end{split}
\end{equation}
where $z^l_x$ is the latent representation of $x$ at $l$-th layer, $\mathbb{N}_u$ and $\mathbb{N}_i$ respectively denote the first-hop neighbors of user $u$ and item $i$.
The enhanced representation of user $u$ is derived as:
$z'_u = \frac{1}{L}\sum_{l=0}^L z^l_u$ with $L$ is the number of graph propagation layers.\\
\textbf{Disentangled Self-Attention Enhance User Preference.} In this phase, \mName{} endeavors to disentangle each user preference into different chunks, which can reflect various aspects of user interests. Specifically, the user latent representation first be disentangled into N-chunks: $z^{(1)}_u \rightarrow z^{(N)}_u \in \mathbb{R}^{d/N}$, $z'_u = \left( z^{(1)}_u \oplus z^{(2)}_u \oplus \dots \oplus z^{(N)}_u \right)$, where $\oplus$ denotes the concat operation. 
Then \mName{} utilizes Multi-Head Self-Attention module~\cite{vaswani2017attention} to sharpen the user latent vector. Each chunk is projected into attention space, formulated as:
\begin{equation}
\begin{split}
   x^{(i,h)}_u &= \eta^{(h)}_{x}(z_u^{(i)}) \in \mathbb{R}^{d/N},\\
   \mathbf{X}^{(h)}_u &= \left[x^{(i,h)}_u\right]_{i=1 } ^{N} \in \mathbb{R}^{N \times d/N},
\end{split}
\end{equation}
where $x^{(i, h)} \in \{q, k, v\}$ stands for query, key, value vectors of user $u$ in $h$-th head and $\eta_x: \mathbb{R}^{d/N} \rightarrow \mathbb{R}^{d/N}$ is a projected layer. The sharpened user's aspect latent feature is derived as:
\begin{equation}
    \bar{z}^{(i, h)}_u = \delta \left( \frac{q_u^{(i,h)} \cdot \mathbf{K}_u^{(j,h)T}}{\sqrt{d/N}} \right) \cdot \mathbf{V}_u^{(j,h)T} \in \mathbb{R}^{d/N},
\end{equation}
where $\delta(\cdot)$ denotes the softmax function. The personalized signal of user $u$: $z^*_u$ is inferred as follows:
\begin{equation}
\begin{split}
    \Tilde{z}_u^{(i)} &= \gamma(\bar{z}_u^{(i, 1)} \oplus \bar{z}_u^{(i, 2)} \oplus \dots \oplus \bar{z}_u^{(i, H)}), \\
    z^*_u &= (\Tilde{z}_u^{(1)} \oplus \Tilde{z}_u^{(2)} \oplus \dots \oplus \Tilde{z}_u^{(N)}),
\end{split}
\end{equation}
where $\gamma : \mathbb{R}^{H \times d/N} \rightarrow \mathbb{R}^{d/N}$ is a feed-forward network and $H$ is the number of attention head.


\subsection{Guided Diffusion Model}

\begin{table*}[t]
\centering
\scalebox{0.93}{
\begin{tabular}{l|cccc|cccc|cccc}
\toprule
Dataset  & \multicolumn{4}{c|}{Youshu}       & \multicolumn{4}{c|}{iFashion}     & \multicolumn{4}{c}{Meal}          \\
Metric    & R@10& R@20& N@10& N@20& R@10& R@20& N@10& N@20& R@10& R@20& N@10& N@20\\
\midrule
CrossCBR~\cite{ma2022crosscbr} &        0.0000&        0.0000&        0.0000&        0.0000&        0.0000&        0.0000&        0.0000&        0.0000&        0.0000&        0.0022&        0.0000&        0.0005\\
 MultiCBR~\cite{ma2024multicbr}& $0.0007$& $0.0025$& $0.0004$& $0.0011$& $0.0074$& $0.0151$& $0.0056$& $0.0088$& $0.0022$& $0.0102$& $0.0028$&$0.0054$\\

CVAE~\cite{sohn2015learning}&        $0.0059$&        $0.0084$&        $0.0040$&        $0.0047$&        $0.0005$&        $0.0011$&        $0.0004$&        $0.0006$&        $0.0045$&        $0.0084$&        $0.0105$&        $0.0111$\\
CLCRec~\cite{wei2021contrastive}   &        $0.0045$&$0.0137$&        $0.0016$&$0.0087$&        $0.0026$&$0.0053$&        $0.0022$&$0.0043$&        $0.0038$&        $0.0064$&        $0.0017$&        $0.0024$\\
CoHeat~\cite{jeon2024cold}   & \underline{$0.0064$}& \underline{$0.0183$}& \underline{$0.0045$}& \underline{$0.0105$}& \underline{$0.0081$}& \underline{$0.0170$}& \underline{$0.0060$}& \underline{$0.0096$}& \underline{$0.1123$}& \underline{$0.1650$}& \underline{$0.0923$}& \underline{$0.1084$}\\
\midrule
\textbf{DiSCo}& $\textbf{0.0303}^\dag$& $\textbf{0.0526}^\dag$& $\textbf{0.0213}^\dag$& $\textbf{0.0283}^\dag$& $\textbf{0.0971}^\dag$& $\textbf{0.1286}^\dag$& $\textbf{0.1094}^\dag$& $\textbf{0.1205}^\dag$& $\textbf{0.2101}^\dag$& $\textbf{0.2838}^\dag$& $\textbf{0.1712}^\dag$& $\textbf{0.1961}^\dag$\\
Imp(\%)  & 373.4 & 187.4 & 373.3 & 169.5 & 1098.7 & 656.4 & 1723.3 & 1155.2 & 87.1 & 72.0 & 85.4 & 80.9 \\
\bottomrule
\end{tabular}
}
\caption{Comparison between \mName{} and baselines on cold-bundles setting, $\dag$ indicates significant improvement with $p < 0.05$.}
\label{compare}
\vspace{-20px}
\end{table*}

For each user $u$, we first sample a bundle represented in distribution space from their past interactions set $\mathcal{X}_u^*$ as $b^{(0)}_u$. Note that $b^{(0)}_u \in \mathbb{R}^{|\mathcal{I}|}$ is represented as an item distribution.\\
\textbf{Diffusion Process.} During the forward process, Gaussian noise is added to the data according to a scheduler $\beta_1, \dots,  \beta_T$ to slowly destroy the original distribution by an approximate posterior function $q(b^{(1:T)}_u|b^{(0)}_u)$.
The inner items of bundle $b^{(0)}$ distribution at timestep $t$ can be achieved directly from $b^{(0)}$ through:
\begin{equation}
    q(b^{(t)}_u|b^{(0)}_u) = \mathcal{N}\left(\sqrt{\bar{\alpha_t}} b^{(0)}_u;(1-\bar{\alpha_t})\mathbf{I}\right),
\end{equation}
where $\mathcal{N}(\mu, \sigma^2)$ denotes normal distribution with mean $\mu$ and variance $\sigma^2$, $\alpha_t = 1 - \beta_t$ and $\bar{\alpha}_t = \prod^t_{s=1} \alpha_s$.\\
\textbf{Reverse Process.} 
The true reverse transition given $b_u^{(0)}$ can be easily formulated according to~\cite{ho2020denoising} as:
\begin{equation}
\begin{split}
    q(b_u^{(t-1)}|b_u^{(0)}, b_u^{(t)}) &= \mathcal{N}\left(\Tilde{\mu}_t(b_u^{(0)}, b_u^{(t)}); \Tilde{\beta}_t\mathbf{I}\right), \\
    \Tilde{\mu}_t(b_u^{(0)}, b_u^{(t)}) &= \frac{\sqrt{\bar{\alpha}_{t-1}}\beta_t}{1-\bar{\alpha}_t} b_u^{(0)} + \frac{\sqrt{\alpha_t}(1-\bar{\alpha}_{t-1})}{1-\bar{\alpha}_t}b_u^{(t)},
\end{split}
\label{true_reverse}
\end{equation}
where $\Tilde{\beta}_t = \frac{1-\bar{\alpha}_{t-1}}{1-\bar{\alpha}_t}\beta_t$. The issue arises when inferring the original distribution without $b_u^{(0)}$, which is undefined. Hence, we estimate the true reverse function using a neural approximator $p_\theta$.
The denoised bundle representation distribution can be obtained using a parameterized reverse transition $p_\theta(b^{(t-1)}_u|b^{(t)}_u)$ as:
\begin{equation}
    p_\theta(b^{(t-1)}_u|b^{(t)}_u) = \mathcal{N}\left(\mu_\theta(b^{(t)}_u,i_u,z^*_u,t); \sigma_t^2 \mathbf{I}\right),
\end{equation}
where $\mu_\theta$ is a neural network parameterized by $\theta$ and the variance $\sigma_t^2  =  \Tilde{\beta}_t$, the same as the true reverse transition~[\ref{true_reverse}].
The architecture of $\mu_\theta$ is described as follows:
\begin{equation}
    \mu_\theta(b^{(t)}_u,i_u,z^*_u,t) = \epsilon_{1} \left(\epsilon_{2}(b^{(t)}_u) \oplus z^*_u\right) + i_u,
\end{equation}
where $\epsilon_{1}: \mathbb{R}^{2d} \rightarrow \mathbb{R}^{|\mathcal{I}|}, \epsilon_{2}: \mathbb{R}^{|\mathcal{I}|} \rightarrow \mathbb{R}^d$ are two MLPs and $i_u$ is the past item interacted distribution of user $u$.

\subsection{Optimization}
\textbf{Simplified Training Objective.} Efficient training is possible by optimizing random term of the full loss function~\cite{ho2020denoising}. Hence, we optimize a simplified version of original term $\mathcal{L}^{(t-1)}$ as:
\begin{equation}
    \mathcal{L}^{(t-1)}_R = \mathbb{E}_{t, u\sim \mathcal{U}, b_u^{(0)} \sim \mathcal{X}_u^*} \left[ \|\Tilde{\mu}_t(b^{(t)}_u,b^{(0)}_u) - \mu_\theta(b^{(t)}_u,t)\|^2 \right],
\end{equation}
where $t \sim Uniform(\{1, 2, \dots, T\})$. The ranking score between user $u$ and bundle $b$ is calculated as follow:
\begin{equation}
    y_{u,b} = b'^{(0)}_u \cdot \Tilde{b}^T,
\end{equation}
where $b'^{(t-1)}_u = \mu_\theta(b^{(t)}_u,t)$ and $b'^{(0)}_u$ is the generated bundle for user $u$, $\Tilde{b}$ is the binary vector of pre-defined bundle $b \in \mathcal{B}$. 
The objective function $\mathcal{L}_R^{(t-1)}$ minimizes the difference between generated bundles and interacted bundles, which boosts the interaction probability between user $u$ and positive bundle $b \sim \mathcal{X}^*_u$.\\
\textbf{Regularization Loss.} To distill the user preference from user-item interactions to the generative process, we adopt a Kullback-Leibler objective term $\mathcal{L}_{KL}$. 
Furthermore, KL loss acts like a regularization term, and the diffusion model can generate a more diverse bundle space rather than merely focus on reconstructing positive bundles.\\
\begin{equation}
\begin{split}
    \mathcal{L}^{(t-1)}_{KL} &= \mathbb{E}_{t, u \sim \mathcal{U}} \left[D_{KL} \left(\delta(i_u)~||~\delta(\mu_\theta(b^{(t)}_u,i_u,z^*_u,t)) \right) \right] \\
                     &= \mathbb{E}_{t, u \sim \mathcal{U}} \left[ \sum \delta(i_u) log\left(\frac{\delta(i_u)}                             {\delta(\mu_\theta(b^{(t)}_u,i_u,z^*_u,t))}\right) \right],
\end{split}    
\end{equation}
The final objective function combines two aforementioned terms defined as follows:
\begin{equation}
    \mathcal{L}^{(t-1)} = \mathcal{L}^{(t-1)}_R + \mathcal{L}_{KL}^{(t-1)} + \tau \| \theta \|^2_2,
\end{equation}
where $\tau$ is a hyper-parameter to control the regularization term.

\begin{figure}[t!]
    \centering
    \begin{subfigure}[b]{0.25\textwidth}
        \centering
        \includegraphics[height=1.2in]{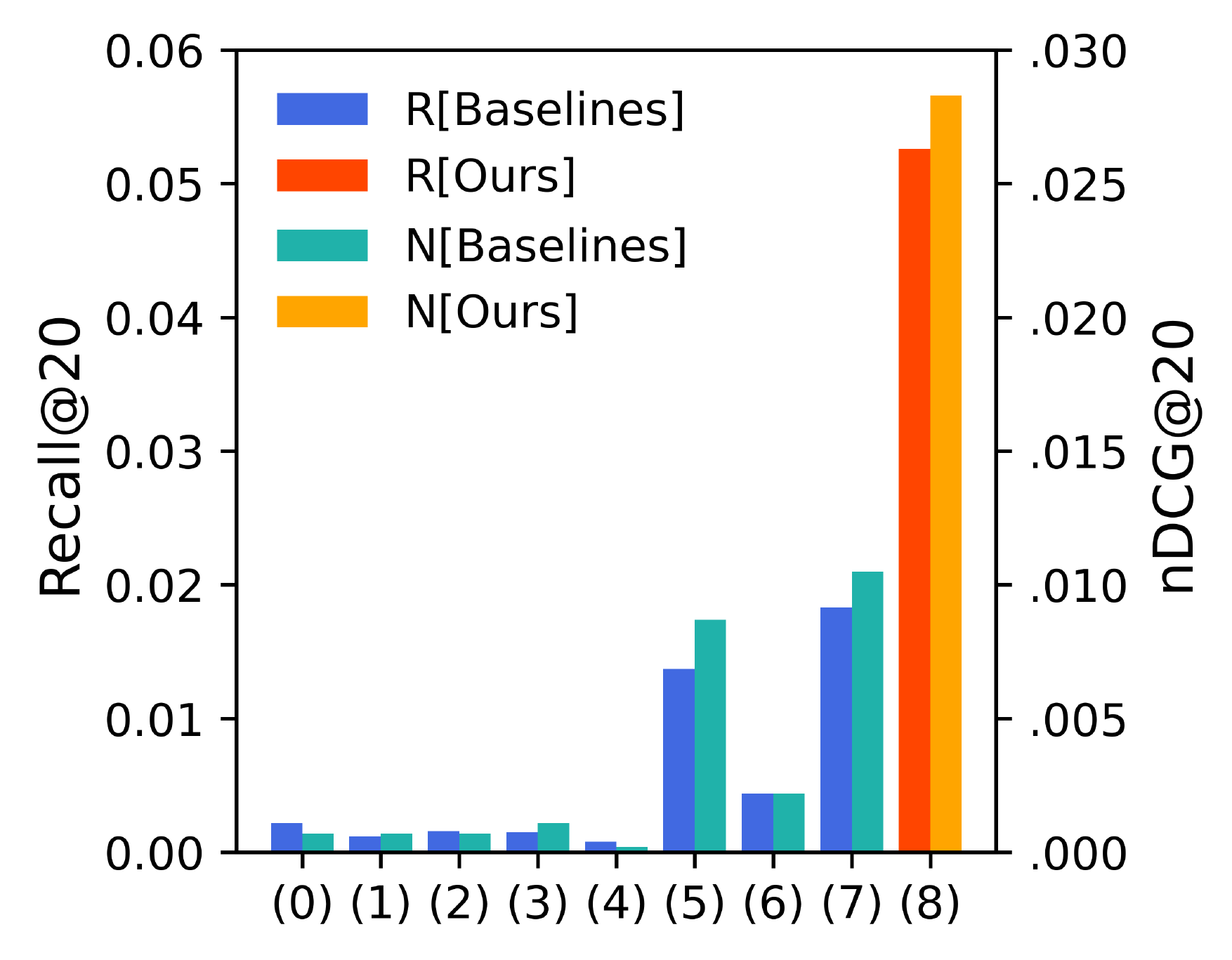}
        \vspace{-5px}
        \caption{Youshu}
    \end{subfigure}%
    ~\hspace{-15px}
    \begin{subfigure}[b]{0.25\textwidth}
        \centering
        \includegraphics[height=1.2in]{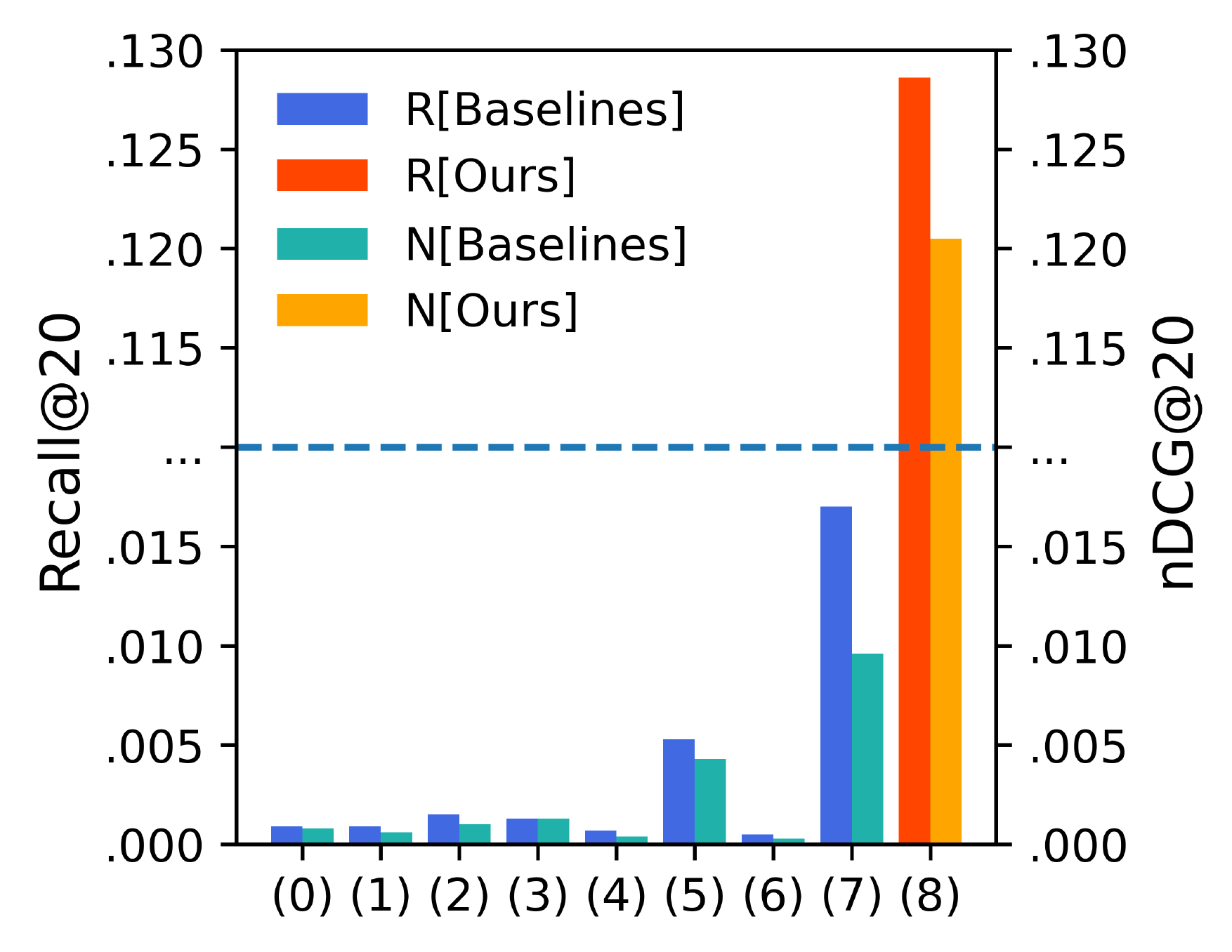}
        \vspace{-5px}
        \caption{iFashion}
    \end{subfigure}
    \vspace{-20px}
    \caption{Comparison with 
        $\textbf{DropoutNet}^{(0)}$\cite{volkovs2017dropoutnet}, 
        $\textbf{CB2CF}^{(1)}$\cite{barkan2019cb2cf}, 
        $\textbf{Heater}^{(2)}$\cite{zhu2020recommendation}, 
        $\textbf{GAR-CF}^{(3)}$\cite{chen2022generative}, 
        $\textbf{CVAR}^{(4)}$\cite{zhao2022improving}, 
        $\textbf{CLCRec}^{(5)}$\cite{wei2021contrastive}, 
        $\textbf{CCFCRec}^{(6)}$\cite{zhou2023contrastive}, 
        $\textbf{CoHeat}^{(7)}$\cite{jeon2024cold} 
        on Youshu and iFashion.}
    \label{cold_baselines}
    \vspace{-15px}
    \Description{Comparison between \mName{} and cold-start recommendation models.}
\end{figure}

\section{Experiments and Results}



\subsection{Experimental Setup}
\textbf{Datasets.} Our experiments are conducted on three datasets: Youshu, iFashion~\cite{ma2022crosscbr} and Meal\footnote{https://github.com/WUT-IDEA/MealRec}. With cold-bundles setting, we divide the bundle set into train, validation, test with ratio $7:1:2$. \\ 
\textbf{Baselines.} 
We compare \mName{} with several approaches, including cold-start recommendation models:
~\cite{zhu2020recommendation, volkovs2017dropoutnet, barkan2019cb2cf, chen2022generative, zhao2022improving, wei2021contrastive, zhou2023contrastive}; bundle recommendation: CrossCBR~\cite{ma2022crosscbr}, MultiCBR~\cite{ma2024multicbr}; along with its cold variant CoHeat~\cite{jeon2024cold} and a conditional generative model CVAE~\cite{sohn2015learning}.
Most baselines results are conducted using official author's implementations with default settings, others are reported in \cite{jeon2024cold}.\\
\textbf{Evaluation.} Following to~\cite{zhu2020recommendation, jeon2024cold}, we rank all bundle then recommend top-$K$ bundle to user. 
We report the averaged result across five different seeds on Recall (R@K) and nDCG (N@K) metrics~\cite{jeon2024cold}.\\
\textbf{Hyper-parameters.} Our model is implemented using Flax model's weights are initialized follows Xavier initialization, the embedding and hidden dimension are tuned in range $\{64, 128\}$, \mName{} is optimized using Adam optimizer. For all datasets the learning rate is set to $1e-3$ with batch size $2048$, number of diffusion step $T=100$, and $\beta_1=1e-5, \beta_T=0.2$. The number of LightGCN and Disentangled-Self-Attention layer is set to 2, $H=2$, N-chunks$=4$, $\tau = 1e-5$.

\subsection{Performance Comparison with Baselines}

The performance of \mName{} outperforms the previous SOTA method by a large margin as shown in Table~\ref{compare}.  Normal bundle recommendation methods~\cite{ma2022crosscbr, ma2024multicbr} do not work for cold-setting due to they rely on user-bundle interactions to update latent embedding which is even sparser in cold-start bundle recommendation. Compare to cold-start bundle variant CoHeat, \mName{} achieves $1155\%$ better nDCG@20 on iFashion and the improvements are still consistent across all datasets.
Additionally, compared with item cold-start recommendation models, our approach outperforms all comparative models by a large margin as shown in Fig~\ref{cold_baselines}. \mName{} obtains $\times 3$ times better performance than second best cold-start recommendation method (CLCRec) on Youshu, similar improvements are observed on iFashion and Meal. The superiority of \mName{} compared to CF-based methods~\cite{ma2022crosscbr, ma2024multicbr} is that \mName{} doesn't rely on poorly update latent space embeddings, our method generate a probabilistic bundle for user, which doesn't rely much on observed interactions between user and bundle, therefore work in every scenarios.

\begin{table}[t]
\scalebox{0.9}{
    \begin{tabular}{l|cc|cc|cc}
    \toprule
    Dataset   & \multicolumn{2}{c|}{Youshu} & \multicolumn{2}{c|}{iFashion} & \multicolumn{2}{c}{Meal} \\
    Metric    & R@20& N@20& R@20& N@20& R@20& N@20\\
    \midrule
    \textit{w/o Graph}&              0.0523&             0.0279&               0.1256&              0.1174&             0.2823&            0.1956\\
    \textit{w/o Disen}& 0.0483       & 0.0263      &               0.1126&              0.1002& 0.2763      & 0.1937     \\
    \textit{w/o KL}& 0.0265       & 0.0174      &               0.0039&              0.0029&             0.0039&            0.0023\\
    \midrule
    \textbf{DiSCo}& \textbf{0.0526}& \textbf{0.0283}&               \textbf{0.1286}&              \textbf{0.1205}& \textbf{0.2838}      & \textbf{0.1961}    \\ \bottomrule
    \end{tabular}
}
\Description{ablation sduty}
\caption{Ablation study on \mName{}}
\label{abl}
\vspace{-15px}
\end{table}

\begin{figure}[t]
    \centering
    \begin{subfigure}[b]{0.25\textwidth}
        \centering
        \includegraphics[height=1.2in]{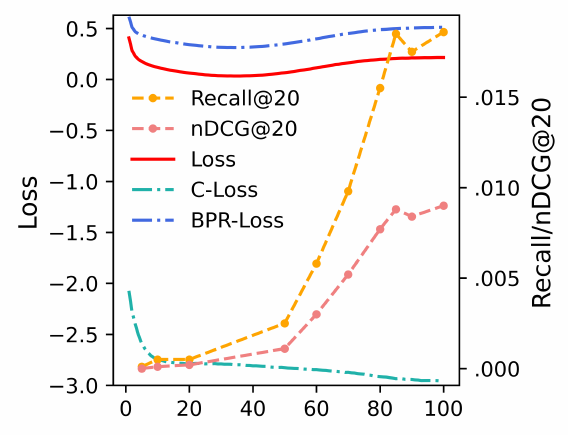}
        \vspace{-5px}
        \caption{CoHeat}
    \end{subfigure}%
    ~\hspace{-15px}
    \begin{subfigure}[b]{0.25\textwidth}
        \centering
        \includegraphics[height=1.2in]{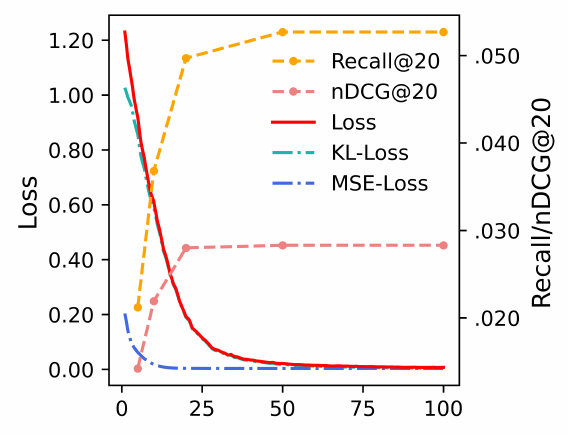}
        \vspace{-5px}
        \caption{\mName{}}
    \end{subfigure}
    \vspace{-20px}
    \caption{Training stability of CoHeat and DisCo on Youshu.}
    \label{stab}
    \vspace{-10px}
    \Description{Training stability comparison between CoHeat and \mName{}.}
\end{figure}

\subsection{Effectiveness of Components}

\textbf{Ablation study.} 
Table~\ref{abl} shows the ablation study on all benchmark dataset of \mName{} and its 3 variants namely: \textit{w/o Graph}, \textit{w/o Disen}, \textit{w/o KL} on cold-bundle setting. In these experiments \textit{w/o Graph} removes the User-Item graph and LightGCN propagation, the user initialized preference vector is directly propagate through Disentangled Self-Attention module. For \textit{w/o Disen}, we pass the user graph-enhanced latent representation directly to the reverse function $p_\theta$. \textit{w/o KL} removes KL-Divergence loss from the model. As the results the performances of all ablated variants consistently drop compared to \mName{}. \textit{w/o KL} shows severe drop in result in all three datasets verify the importance of minimizing the difference between user's bundle and user's item preference.\\
\textbf{Stability.} We conduct an additional experiment to verify the stability of \mName{} as shown in Fig~\ref{stab}. CoHeat achieve its peak performance when it's loss function start to increase while the performance of \mName{} increase consistently while the objective loss function drop.
The optimal point while optimizing the objective function come with the peak performance of \mName{}.


\section{Conclusion}
In this work we focus on cold-start bundle recommendation problem where common latent-based methods fail. Our proposed approach \mName{} can tackle the problem of cold-start setting about representing cold bundles by generating a personalized probabilistic bundle for each user instead of using latent feature by utilizing diffusion backbone guided by user latent preference and a regularization loss that help avoiding bias while learning on the interacted bundle and inject the item preference into the generative process. Extensive experiments show the effectiveness of our method achieving new SOTA results on all benchmark datasets in cold setting.


\bibliographystyle{ACM-Reference-Format}
\bibliography{main}


\end{document}